\documentclass[twocolumn,prl,aps,showpacs]{revtex4}
\usepackage{graphicx}

\begin{document}

\newcommand{\snn}{\sqrt{s_{NN}}}
\newcommand{\seff}{\sqrt{s_{\rm eff}}}
\newcommand{\s}{\sqrt{s}}
\newcommand{\pp}{p+p}
\newcommand{\pbarp}{\overline{p}+p}
\newcommand{\qbarq}{\overline{q}q}
\newcommand{\epem}{e^+e^-}

\newcommand{\nhit}{N_{hit}}
\newcommand{\npp}{n_{pp}}
\newcommand{\nch}{N_{ch}}
\newcommand{\avench}{\langle\nch\rangle}
\newcommand{\np}{N_{part}}
\newcommand{\ns}{N_{spec}}
\newcommand{\ntot}{\langle\nch\rangle}
\newcommand{\avenp}{\langle\np\rangle}
\newcommand{\npB}{N_{part}^B}
\newcommand{\nc}{N_{coll}}
\newcommand{\avenc}{\langle\nc\rangle}
\newcommand{\half}{\frac{1}{2}}
\newcommand{\halfnp}{\langle\np/2\rangle}
\newcommand{\etap}{\eta^{\prime}}
\newcommand{\as}{\alpha_{s}(s)}
\newcommand{\etazero}{\eta = 0}
\newcommand{\etaone}{|\eta| < 1}
\newcommand{\dndeta}{d\nch/d\eta}
\newcommand{\dndetazero}{\dndeta|_{\etazero}}
\newcommand{\dndetaone}{\dndeta|_{\etaone}}
\newcommand{\dndetanp}{\dndeta / \halfnp}
\newcommand{\dndetaonp}{\dndeta / \np}
\newcommand{\dndetazeronp}{\dndetazero / \halfnp}
\newcommand{\dndetaonenp}{\dndetaone / \halfnp}
\newcommand{\ratio}{\ntot/\halfnp}
\newcommand{\nee}{N_{\epem}}
\newcommand{\nhh}{N_{hh}}
\newcommand{\nubar}{\overline{\nu}}
\newcommand{\rpcnpart}{R^{\np}_{PC}}
\title{Charged-Particle Pseudorapidity Distributions in Au+Au Collisions at $\snn=62.4$ GeV}

\author{
B.B.Back$^1$,
M.D.Baker$^2$,
M.Ballintijn$^4$,
D.S.Barton$^2$,
R.R.Betts$^6$,
A.A.Bickley$^7$,
R.Bindel$^7$,
W.Busza$^4$,
A.Carroll$^2$,
Z.Chai$^2$,
M.P.Decowski$^4$,
E.Garc\'{\i}a$^6$,
T.Gburek$^3$,
N.George$^2$,
K.Gulbrandsen$^4$,
C.Halliwell$^6$,
J.Hamblen$^8$,
M.Hauer$^2$,
C.Henderson$^4$,
D.J.Hofman$^6$,
R.S.Hollis$^6$,
R.Ho\l y\'{n}ski$^3$,
B.Holzman$^2$,
A.Iordanova$^6$,
E.Johnson$^8$,
J.L.Kane$^4$,
N.Khan$^8$,
P.Kulinich$^4$,
C.M.Kuo$^5$,
W.T.Lin$^5$,
S.Manly$^8$,
A.C.Mignerey$^7$,
R.Nouicer$^{2,6}$,
A.Olszewski$^3$,
R.Pak$^2$,
C.Reed$^4$,
C.Roland$^4$,
G.Roland$^4$,
J.Sagerer$^6$,
H.Seals$^2$,
I.Sedykh$^2$,
C.E.Smith$^6$,
M.A.Stankiewicz$^2$,
P.Steinberg$^2$,
G.S.F.Stephans$^4$,
A.Sukhanov$^2$,
M.B.Tonjes$^7$,
A.Trzupek$^3$,
C.Vale$^4$,
G.J.van~Nieuwenhuizen$^4$,
S.S.Vaurynovich$^4$,
R.Verdier$^4$,
G.I.Veres$^4$,
E.Wenger$^4$,
F.L.H.Wolfs$^8$,
B.Wosiek$^3$,
K.Wo\'{z}niak$^3$,
B.Wys\l ouch$^4$\\
\vspace{3mm}
\small
$^1$~Argonne National Laboratory, Argonne, IL 60439-4843, USA\\
$^2$~Brookhaven National Laboratory, Upton, NY 11973-5000, USA\\
$^3$~Institute of Nuclear Physics PAN, Krak\'{o}w, Poland\\
$^4$~Massachusetts Institute of Technology, Cambridge, MA 02139-4307, USA\\
$^5$~National Central University, Chung-Li, Taiwan\\
$^6$~University of Illinois at Chicago, Chicago, IL 60607-7059, USA\\
$^7$~University of Maryland, College Park, MD 20742, USA\\
$^8$~University of Rochester, Rochester, NY 14627, USA\\
}
\date{\today}

\begin{abstract}
The charged-particle pseudorapidity density for Au+Au
collisions at $\snn$=62.4 GeV has been measured over a wide
range of impact parameters and compared to results obtained
at other energies.
As a function of collision energy, the pseudorapidity 
distribution grows systematically both in height and width.  The
mid-rapidity density is found to grow approximately logarithmically 
between AGS energies and the top RHIC energy.  As a function of centrality,
there is an approximate factorization of the centrality dependence of
the mid-rapidity yields and the overall multiplicity scale.  
The new results at $\snn$=62.4 GeV confirm the previously
observed phenomenon of ``extended longitudinal scaling'' in the
pseudorapidity distributions when viewed in the rest frame of one
of the colliding nuclei. 
It is also found that the evolution of the shape of the 
distribution with centrality is energy independent, when viewed
in this reference frame. 
As a function of centrality, the total charged
particle multiplicity scales linearly with the number of participant
pairs as it was observed at other energies.

\end{abstract}

\maketitle

In previous publications the PHOBOS Collaboration 
has presented the full systematic behavior
of inclusive charged particle production in heavy ion collisions over
a large range of collision parameters:
1) $\snn$ from 19.6 GeV to 200 GeV, 2) Pseudorapidities from 
$\eta=-5.4$ to $5.4$, nearly the full solid angle, and 3)
average impact parameter from $\langle b\rangle =3-10.5$ fm, 
corresponding to 50-360 participating nucleons ($\np$)
~\cite{Back:2000gw,Back:2001ae,Back:2001bq,Back:2001xy, Back:2002uc,Back:2002wb,Back:2004dy}.
While these data are already useful as a broad systematic study, 
several non-trivial 
features
have been made manifest
by direct comparisons between the data at different energies
and centralities:
1) a logarithmic increase with $\snn$ in the mid-rapidity particle density~\cite{Back:2001ae},
2) an approximate factorization of the centrality and energy
dependence of the mid-rapidity yields ~\cite{Back:2004dy}
3) the phenomenon of ``limiting fragmentation'' in the forward direction~\cite{Back:2002wb}, 
and 4) a linear ``$\np$-scaling'' of the total particle yield~\cite{Back:2003xk}.
This paper presents for the first time the multiplicity data for
Au+Au collisions at
the most recent RHIC energy of $\snn=62.4$ GeV, corresponding to the top
energy reported by several earlier $\pp$ experiments at the CERN ISR.
With our new data, we can test the previously-found scaling
relationships at an intermediate energy.

\begin{figure}[t]
\begin{center}
\includegraphics[width=7.5cm]{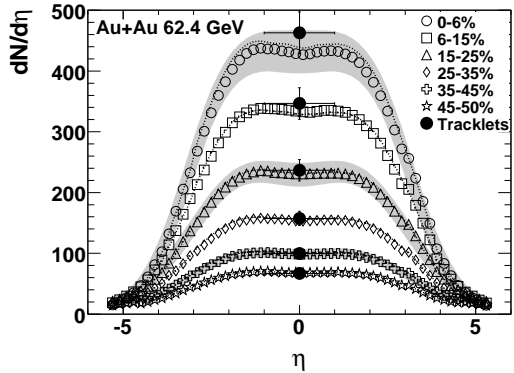}
\end{center}
\caption{
Pseudorapidity distributions $\dndeta$ from Au+Au
collisions at $\snn=62.4$ GeV.  
Open symbols show the
results obtained using the paddle-based
centrality method with 90\% C.L. systematic errors
indicated by grey bands for selected bins.  
The dotted lines show the results obtained using the
Octagon-based centrality method.
The filled circles
show the average value of $\dndetaone$ using the tracklet technique,
with the horizontal bars indicating the interval in $\eta$
over which the averaging is done.
The centrality is denoted by the fraction of the
total inelastic cross section, with smaller numbers corresponding
to more central events.
}
\label{figure1}
\end{figure}

The data were obtained with the PHOBOS detector\cite{Back:2003sr}
during the RHIC 2004 run. In this analysis only
the data taken with the magnetic field switched off are analyzed.
To select events with various ranges of impact parameter,
characterized typically by $\np$, 
we use the particle multiplicity measured in two sets
of ``paddle'' counters, situated at $z=\pm 3.21$ meters from
the nominal interaction point.  These cover a pseudorapidity range
of $3.2<|\eta|<4.5$ with 95\% azimuthal acceptance. 
The overall triggering and event selection
efficiency at 62.4 GeV corresponds to $81 \pm 2 \%$ of the total
Au+Au inelastic cross section, estimated using HIJING 
simulations~\cite{Gyulassy:1994ew}.
We use the Glauber model calculation implemented in HIJING to
estimate $\avenp$ for each centrality 
bin by assuming a monotonic relationship
between $\np$ and the relevant experimental observable.
This procedure has been described in Refs.~\cite{Back:2000gw,Back:2001xy,Back:2002uc}.
It was found that trigger efficiencies are
typically around 100\% for the top 50\% of the total cross section
at energies of $\snn=62.4,130$ and $200$ GeV.
At the lowest energy of $\snn=19.6$ GeV,
an alternative method of centrality determination was 
developed, that uses all
particles detected in the ``Octagon'' silicon detector,
covering $|\eta|<3$, as a measure of the particle 
multiplicity~\cite{Back:2002wb}.
These two methods agree within 3\% at the higher energies 
and have been used as a cross-check in this analysis.
However it should be noted that the two methods give 
values of $\avenp$ that differ by $2\%$
in the most central bin, since this bin is more sensitive to the
fluctuations of the variable used to estimate the centrality.
Thus, we separate the tabulated results (in Table \ref{table1}) 
for that bin, but do not in the other bins when they 
agree to better than $1\%$.

Several methods were used to estimate the charged
particle density in each centrality class.  In the full phase space,
a combination of data from the single-layer 
``Octagon'' ($|\eta|<3$) and ``Ring''
($3<|\eta|<5.4$) detectors were analyzed using two different techniques.
In the ``analog'' method, the
deposited energy in a detector pad is used to estimate the number of
particles traversing the pad after accounting for
orientation of the Si-wafer relative to the interaction point.
The ``digital'' approach treats each
pad as a binary counter and assumes Poisson statistics to estimate
the total occupancy in various regions of pseudorapidity.  These methods
have been discussed in more detail in Refs.~\cite{Back:2001bq} 
and~\cite{Back:2002wb}.
At mid-rapidity, the PHOBOS vertex detector, consisting of two
planes covering $|\eta|<0.92$ over a limited azimuthal range, 
$\Delta \phi \sim 90^{o}$, is used
to count ``tracklets''.  These are two-hit tracks which
point back to the event vertex, providing 
redundancy not present in the single-layer analyses, and thereby
reducing systematic
effects at mid-rapidity.  This method has been described in detail in
Refs.~\cite{Back:2001bq, Back:2002uc, Back:2004dy}.

\begin{figure}[t]
\begin{center}
\includegraphics[width=7.5cm]{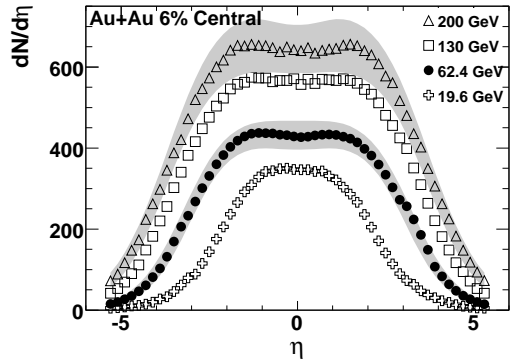}
\end{center}
\caption
{
The pseudorapidity distributions measured in the 6\% most-central
Au+Au collisions at four RHIC energies.  
90\% C.L. systematic errors are shown as grey bands.
\label{figure2}
}
\end{figure}

The $\snn=62.4$ GeV data are shown in Fig.~\ref{figure1} as
a function of collision centrality, determined by the paddle-counter method. 
Results using the Octagon-based centrality method, shown for
each bin by dotted lines, agree very well with those
from the paddle-based method.
Mid-rapidity data from the tracklet method 
also agree well with
the single-Si-layer analysis over the full centrality range studied.  

To place these data in context, Fig.~\ref{figure2} shows
data from the $6\%$ most
central collisions 
in comparison with similar data at
19.6, 130, and 200 GeV 
from Ref.~\cite{Back:2002wb}.  
Increases both in the height and width of the distribution
are observed as a function of increasing energy.  
Already at $\snn=62.4$ GeV, the central ``plateau''
nascent at $\snn=19.6$ GeV is fully developed and grows
in width slowly with increasing energy.  
Of course, the existence of a plateau in $\dndeta$ does not
necessarily imply the existence of a plateau in $dN/dy$ because
of the non-trivial transformation between rapidity and
pseudorapidity ($dy = \beta d \eta$)~\cite{Back:2002wb}.

\begin{figure}[t]
\begin{center}
\includegraphics[width=7.5cm]{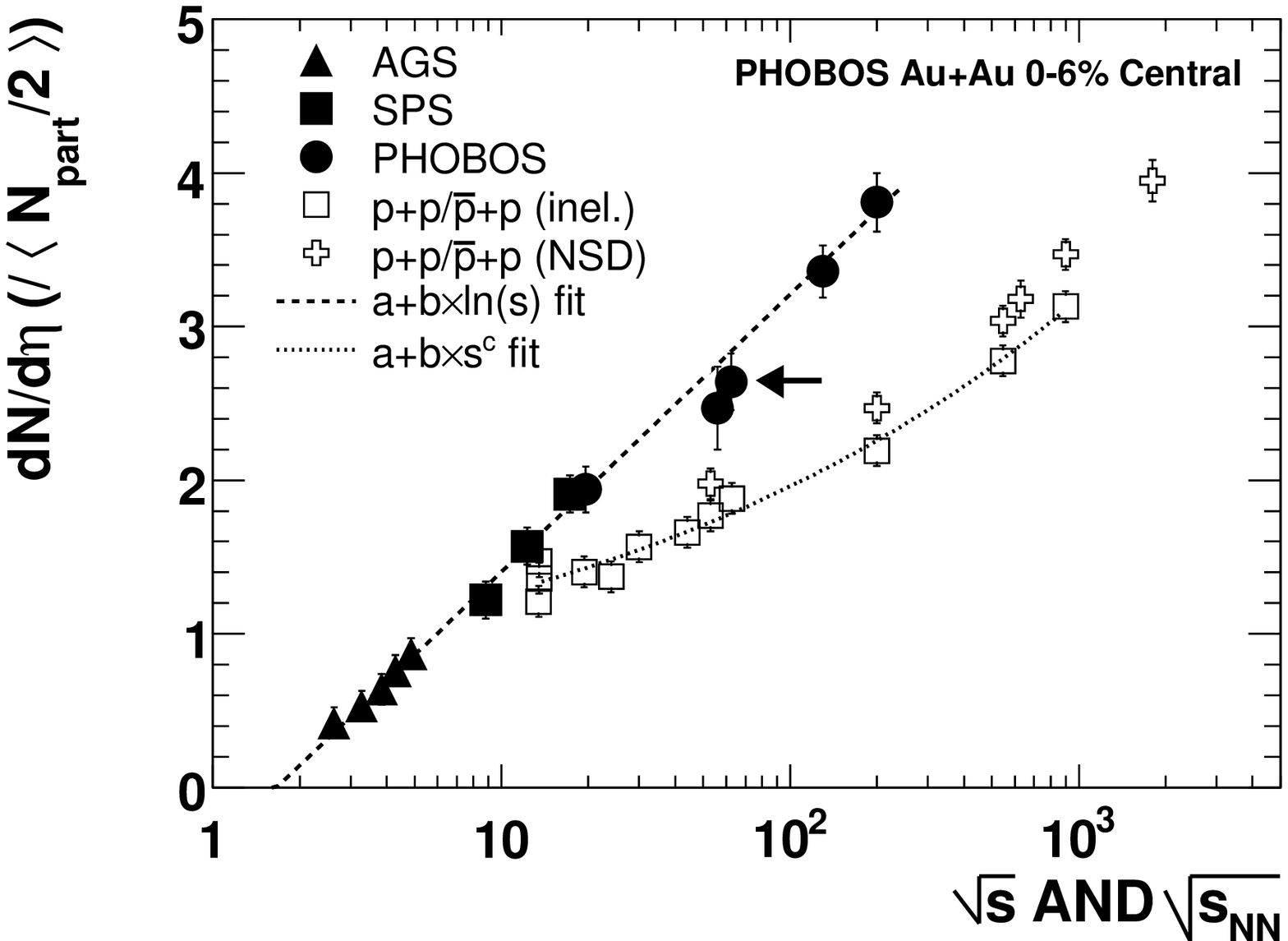}
\end{center}
\caption{
$\dndetaonenp$ shown for Au+Au collisions as a function of energy.  
The PHOBOS
data, averaged over all available measurement techniques,
is compared with lower-energy A+A data as well as a variety
of $\pp$ and $\pbarp$ data.  The thick dashed line is a fit ($a+b\ln(s)$, with $a=-0.40$ and $b=0.39$) to the
$\snn=19.6,130$ and $200$ GeV data points.
The inelastic $\pp$ and $\pbarp$ 
data have been fit by a function $a+bs^{c}$ (with $a=0.35$, $b=0.52$ and $c=0.12$), shown
by a thin dotted line, used only for interpolation.
\label{figure3}
}
\end{figure}

The scaled particle densities near midrapidity ($\dndetaonenp$) for
$\snn=19.6, 130$ and $200$ GeV~\cite{Back:2000gw,Back:2001xy,
Back:2001ae,Back:2002uc,Back:2002wb,Back:2004dy, Back:2004je}
are shown as a function
of collision energy in Fig.~\ref{figure3} for the 6\% most central
events.  Where possible, PHOBOS results from the various measurement
techniques have been averaged at each energy, weighted by the 
inverse square of
the relative error.  At $\snn=62.4$ GeV, this gives
$\dndetaonenp = 2.64 \pm 0.18$.
Data from comparable centralities at lower energies
are also shown, as compiled in Ref.~\cite{Back:2004je}.
While high-statistics data from the three other RHIC 
energies suggested an approximately-logarithmic rise of the particle density,
the low-statistics data point measured at $\snn=56$ GeV was found
to be only barely consistent with a logarithmic fit based only
on the data at $\snn=19.6, 130$ and $200$ GeV (shown as a dashed
line).  The data point at
$\snn=62.4$ GeV (indicated by an arrow in Fig.~\ref{figure3})
falls closer to the fit, and is, within errors, consistent with
the logarithmic rise.
The $A+A$ data are compared to a wide range of $\pp$
and $\pbarp$ data, separately shown for inelastic as
well as non-single diffractive events~\cite{ua5,ua5physrep,Abe:1989td}.  
While these data also
appear to rise logarithmically at higher energies, 
the inclusion of data points below
$\s \sim$ 30 GeV appears to indicate a curvature in the
energy dependence.
To interpolate between the
measured points, they have been fit by a function $a+bs^{c}$, shown
by the thin dashed line.

\begin{figure}[t]
\begin{center}
\includegraphics[width=7.5cm]{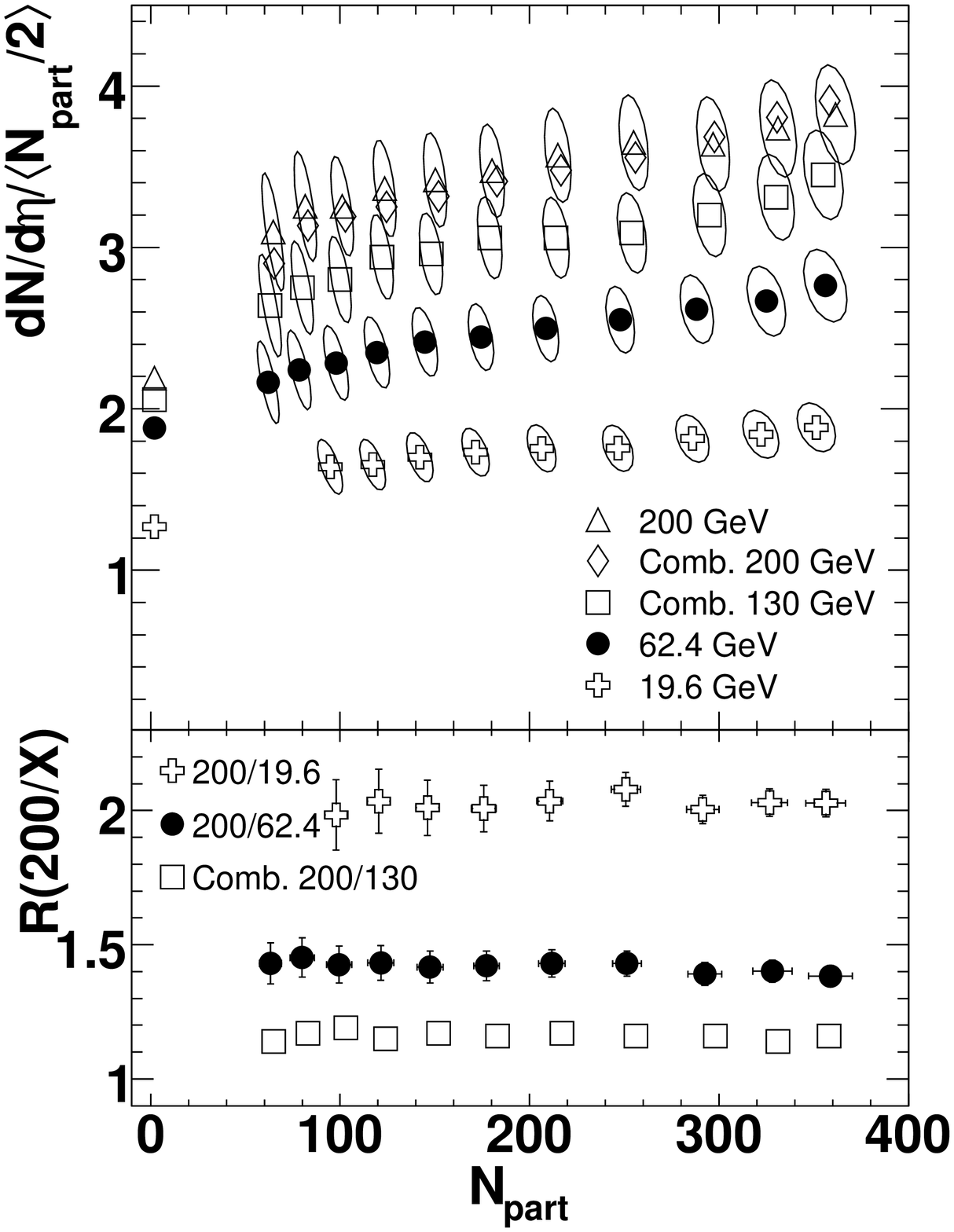}
\end{center}
\caption{
a.) $\dndetaonenp$ measured using the tracklet technique as a function
of $\np$ for four RHIC energies ($\snn=19.6,62.4,130$ and $200$ GeV).
Also shown are inelastic $\pp$ data for $\s=62.4$ and $\s=200$ GeV, 
and interpolated values for $\s=19.6$ GeV (Ref.~\cite{Back:2004dy}) and
$\s=130$ GeV (using the fit shown 
in Fig.~\ref{figure3}).  The systematic errors are shown as
90\% C.L. ellipses (reflecting the trivial correlation in $\dndetaonenp$
and $\np$).
b.) The ratio of the scaled pseudorapidity densities measured at
200 GeV to those measured at $\snn=19.6,62.4$ and $130$ GeV.
\label{figure4}
}
\end{figure}

The centrality dependence of $\dndetaonenp$, measured with the
tracklet technique, is shown in Fig.~\ref{figure4}a, and
tabulated in Table~\ref{table1}.  Data
at $\snn=19.6, 62.4,$ and $200$ GeV have been analyzed using only the
tracklet technique in the PHOBOS Vertex detector (hereafter called
the ``vertex-tracklet'' method) and the Octagon-based centrality
method.
At 130 and 200 GeV, results are also available using a method
(the ``combined'' method) which averages the vertex tracklet results 
with a similar method using the PHOBOS spectrometer~\cite{Back:2002uc},
and using the paddle-based centrality method.
The results are typically compatible
within 2\% over the full centrality range, as can be seen by
direct comparison in Fig.~\ref{figure4}a of the 
vertex tracklet and the combined result for 200 GeV.
The vertex-tracklet method and Octagon-based centrality method 
is used at 62.4 GeV for overall consistency and partial
cancellation of certain systematics in the ratios relative to 19.6
and 200 GeV data.  

The centrality dependence of the mid-rapidity yields has often been
interpreted in a two-component picture of particle production,
with ``soft'' processes scaling with $\np$ and ``hard'' processes
scaling with the number of binary collisions, $\nc$.  As the
beam energy increases, particle production 
from the hard processes, which exceed the number of participant
pairs by a factor of $\sim 5-6$ in central events for $\snn$
ranging from 19.6 to 200 GeV, 
are expected to dominate over that from
soft processes as the minijet cross sections increase~\cite{Gyulassy:1994ew}.
This expectation may be examined by studying the ratio of the yields at
different energies for the same fraction of the total cross section,
shown in Fig.~\ref{figure4}b.  Despite the expected increase
in hard processes with increasing energy, these ratios are
observed to be
constant over the measured centrality range, showing a
``factorization'' of beam energy and collision geometry at midrapidity.
This result extends
the analysis presented in Ref.~\cite{Back:2004dy} and is 
fully compatible with the constant ratio found in comparisons between 
data at $\snn=$ 200 and 130 GeV obtained with the ``combined'' method.
All of these data suggest that while two-component models can fit
the midrapidity data at each energy, they do not have an energy dependence 
characteristic of a growing contribution of hard-processes.

\begin{figure}[t]
\begin{center}
\includegraphics[width=7.5cm]{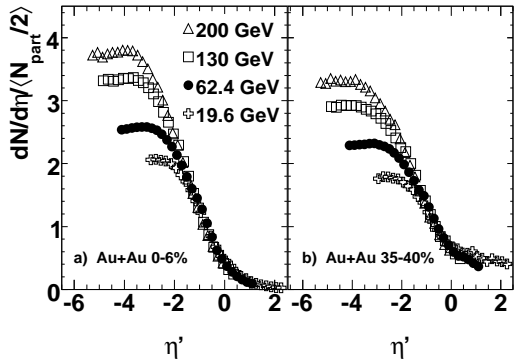}
\end{center}
\caption{
The scaled pseudorapidity density ($\dndetanp$) for two
centrality bins (0-6\% and 35-40\%) and four RHIC energies 
shown in the rest frame of one of the projectiles by using
the variable $\etap=\eta-y_{beam}$.  For clarity, systematic errors
are not shown in this figure.
\label{figure5}
}
\end{figure}

\begin{figure}[b]
\begin{center}
\includegraphics[width=7.5cm]{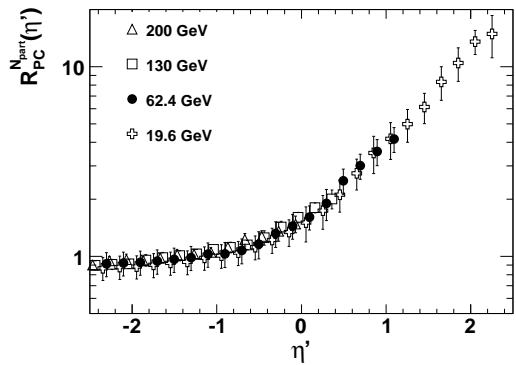}
\end{center}
\caption{
$\rpcnpart$ as a function of $\etap$
comparing the 35-40\% bin to the 0-6\% most central
bin for four different energies.
\label{figure5a}
}
\end{figure}

In a previous PHOBOS publication, the phenomenon of 
``limiting fragmentation'' was observed by comparing Au+Au collisions
at the three RHIC energies for which $4\pi$ data were available~\cite{Back:2002wb}.
This phenomenon, which we also refer to as ``extended longitudinal
scaling''~\cite{Back:2004mr} is simply the invariance of the scaled yields
$\dndetanp$ with beam energy in the reference frame 
of one of the projectiles, i.e. by plotting the
scaled yields with respect to the variable $\etap = \eta - y_{beam}$,
where $y_{beam}$ is the beam rapidity.
The concept of extended longitudinal scaling is expected to
apply to rapidity distributions, but since $\eta\sim y-\ln(p_T/m_T)$
for particles emitted far away from mid-rapidity,
this scaling is also expected to apply to pseudorapidity distributions.
This scaling phenomenon was also found in $d+Au$ collisions 
at RHIC~\cite{Back:2004mr} 
and, surprisingly, also for elliptic flow at
all of the RHIC energies, including $\snn=62.4$ GeV~\cite{Back:2004zg}.
The new multiplicity data at $\snn=62.4$ GeV, 
shown in Figs.~\ref{figure5}a and \ref{figure5}b, 
clearly fit well into the existing pattern.

The data in Fig.~\ref{figure5} suggest that the change in the shape of 
$\dndeta$  away from mid-rapidity is a stronger function of the 
collision geometry than of the beam energy,
when observed in the rest frame of one of the projectiles.
This is shown more clearly in Fig.~\ref{figure5a}, where the
ratio of peripheral to central data scaled by $\np$
\[
\rpcnpart(\etap,35-40\%) = \frac{\np^{0-6\%}}{\np^{35-40\%}}
\frac{\dndeta^{35-40\%}}{\dndeta^{0-6\%}}
\]
is plotted as a function of $\etap$.
The error bars in this figure indicate 90\% C.L. systematic errors.
A similar figure was shown in Ref.~\cite{Back:2002wb}, showing that
the change in shape as a function of centrality is independent of beam
energy when observed in the rest frame of one of the projectiles.
This ratio was also used to compare the
shapes of transverse momentum distributions, $dN/dp_T$,
measured near mid-rapidity
for different energies and centralities, and
a similar invariance with energy was found for each 
centrality bin~\cite{Back:2004ra}, i.e. 
in all of these cases, the centrality and energy dependences
factorize to a large extent.  Such behavior has also been seen
in proton-nucleus collisions at lower energies~\cite{Busza:1991ah}.

\begin{table}[t]
\begin{tabular}{|c|c|c|c|c|c|}
\hline
\multicolumn{2}{|c}{Centrality}&
\multicolumn{2}{|c|}{Fig. 4}&
\multicolumn{2}{c|}{Fig. 7}\\
\hline
Bin & $\np$ & $dN/d\eta$   & $\dndeta/$ & $\nch$ & $\nch/$ \\
    &       & $(|\eta|<1)$ & $\halfnp$  &        & $\halfnp$ \\
\hline
0-3\%& 356$\pm$11 & 492$\pm$36 & 2.76$\pm$0.23 & &  \\
     & 349$\pm$11 &  &  & 2988$\pm$149  & 17.10$\pm$1.02 \\
3-6\%& 325$\pm$10 & 433$\pm$32 & 2.67$\pm$0.22 & 2775$\pm$138 & 17.17$\pm$1.01 \\
6-10\%& 288$\pm$ 9 & 377$\pm$28 & 2.62$\pm$0.21 & 2489$\pm$124 & 17.29$\pm$1.01 \\
10-15\%& 248$\pm$ 8 & 316$\pm$23 & 2.55$\pm$0.21 & 2120$\pm$106 & 17.13$\pm$1.00 \\
15-20\%& 209$\pm$ 7 & 260$\pm$19 & 2.50$\pm$0.20 & 1777$\pm$88 & 17.03$\pm$1.02 \\
20-25\%& 174$\pm$ 7 & 212$\pm$15 & 2.44$\pm$0.21 & 1485$\pm$74 & 17.07$\pm$1.08 \\
25-30\%& 145$\pm$ 7 & 174$\pm$13 & 2.41$\pm$0.21 & 1236$\pm$61 & 17.03$\pm$1.17 \\
30-35\%& 119$\pm$ 7 & 140$\pm$10 & 2.35$\pm$0.22 & 1027$\pm$51 & 17.15$\pm$1.30 \\
35-40\%&  98$\pm$ 7 & 111$\pm$ 8 & 2.28$\pm$0.23 & 840$\pm$42 & 17.17$\pm$1.45 \\
40-45\%&  78$\pm$ 6 &  87$\pm$ 6 & 2.24$\pm$0.25 & 679$\pm$33 & 17.30$\pm$1.63 \\
45-50\%&  62$\pm$ 6 &  67$\pm$ 5 & 2.16$\pm$0.26 & 532$\pm$26 & 17.16$\pm$1.80 \\
\hline
\end{tabular}
\caption{
Data for Au+Au collisions at $\snn=62.4$ GeV including 
$\dndetaone$, $\dndetaonenp$ as shown in Fig. 4
and $\ntot$ and $\ratio$ as shown in Fig. 7.  
The difference in $\np$ in the $0-3\%$ bin is explained in the text.
\label{table1}
}
\end{table}

\begin{figure}[t]
\begin{center}
\includegraphics[width=7.cm]{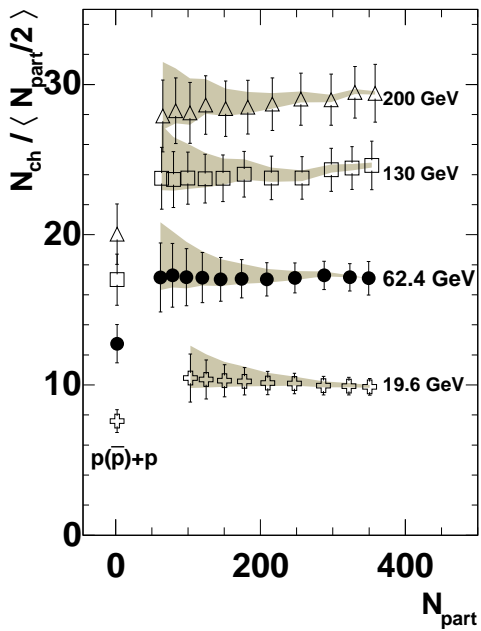}
\end{center}
\caption{
$\ratio$, obtained by extrapolating the data at
each energy into the unmeasured region, as a function of
centrality.  The 90\% C.L. uncertainty on $\nch$ and $\np$ have been
combined into the error bars, while the 90\% C.L. uncertainty on the
extrapolation procedure is indicated by a grey band.
Inelastic $\pp$ and $\pbarp$ data, interpolated using a power-law fit, 
are shown at $\np=2$.
}
\label{figure6}
\end{figure}

As observed previously, the centrality dependence
of the limiting curve has the interesting property that the decrease
in the scaled particle density, $\dndetanp$,
at mid-rapidity when moving from central to peripheral events
is correlated with the increase at forward rapidities. 
Although some of
the particles with $\etap>0$ may be attributed to emission from spectators,
the systematic change of the slope leads to an approximately
constant total multiplicity.  Using the method outlined in
Ref.~\cite{Back:2003xk}, which combines analytic fits of the 
measured region with estimates of the unmeasured yields using
the lower-energy data, the total charged-particle multiplicity
extrapolated to $4\pi$ has been calculated as a function of centrality,
as shown in Fig.~\ref{figure6} and tabulated in 
Table~\ref{table1}.
The uncertainty from the extrapolation procedure itself is indicated
by the grey bands.  As at the other RHIC energies~\cite{Back:2003xk}, 
the data at $\snn=62.4$ GeV shows an approximately linear relationship between
$\nch$ and $\np$.  This persistence of 
``wounded nucleon scaling''~\cite{elias,wounded} 
has not been fully explained for heavy ion collisions, especially
since the multiplicity is evidently not a simple 
multiplication of nucleon-nucleon multiplicity by $\np/2$.

It is an interesting question whether or not the various scaling
behaviors discussed, such as the factorization of energy and
geometry at midrapidity (Fig. \ref{figure4}) and the similar 
factorization of the distributions in $\etap$ (Fig. \ref{figure5a}), 
should be considered independent phenomena.  
Already, the
$\np$ scaling shown in Fig. \ref{figure6} suggests that modifications
to particle production at forward rapidities are strongly correlated
with compensating changes at mid-rapidity.  
If, in fact, the
pseudorapidity distribution at each energy deviates from the limiting
curve at around $\eta \sim 1.5-2$ by flattening out, 
and this deviation is centrality 
independent (a statement which is broadly consistent with the available
data, except perhaps at the lowest RHIC energy), then the
factorization of energy and geometry at mid-rapidity follows naturally
as a consequence of the centrality dependence of the energy-independent
limiting curve.
The same condition also relates the approximately logarithmic energy 
dependence (Fig. \ref{figure3}) to the shape of the limiting curve
in central events.
Of course, empirical observations like this do not {\it explain} 
why these relationships between the various regions of phase space
hold, but rather point to issues that need to
be addressed in the global understanding of heavy ion collisions.

In summary,
the charged-particle pseudorapidity density has been measured by PHOBOS  
for Au+Au collisions at $\snn$= 62.4 GeV, matching
the top ISR energy.  As a function of collision energy, the pseudorapidity 
distribution grows systematically both in height and width.  The
mid-rapidity density is found to grow approximately logarithmically 
between AGS energies and the top RHIC energy.  As a function of centrality,
there is an approximate factorization of the centrality dependence of
the mid-rapidity yields and the energy-dependent overall multiplicity scale. 
The phenomenon of ``extended longitudinal scaling'' (also known as 
``limiting fragmentation'')
is clearly present in the 62.4 GeV data.
Finally, a relatively-small extrapolation of the measured
yields to 4$\pi$ allows the
extraction of the total charged-particle multiplicity.
As at the other RHIC energies, $\nch$ is found to
scale approximately linearly with the number of participants
over the range of collision centralities studied.

This work was partially supported by U.S. DOE grants 
DE-AC02-98CH10886,
DE-FG02-93ER40802, 
DE-FC02-94ER40818,  
DE-FG02-94ER40865, 
DE-FG02-99ER41099, and
W-31-109-ENG-38, by U.S. 
NSF grants 9603486, 
0072204,            
and 0245011,        
by Polish KBN grant 1-P03B-062-27(2004-2007),
by NSC of Taiwan Contract NSC 89-2112-M-008-024, and
by Hungarian OTKA grant (F 049823).


\begin{references}
\bibitem{Back:2000gw} B.~B.~Back {\it et al.}, Phys.\ Rev.\ Lett.\  {\bf 85}, 3100 (2000). 
\bibitem{Back:2001ae} B.~B.~Back {\it et al.}, Phys.\ Rev.\ Lett.\  {\bf 88}, 022302 (2002). 
\bibitem{Back:2001bq} B.~B.~Back {\it et al.}, Phys.\ Rev.\ Lett.\  {\bf 87}, 102303 (2001). 
\bibitem{Back:2001xy} B.~B.~Back {\it et al.}, Phys.\ Rev.\ C {\bf 65}, 031901 (2002).
\bibitem{Back:2002uc} B.~B.~Back {\it et al.}, Phys.\ Rev.\ C {\bf 65}, 061901 (2002).
\bibitem{Back:2002wb} B.~B.~Back {\it et al.}, Phys.\ Rev.\ Lett.\  {\bf 91}, 052303 (2003).
\bibitem{Back:2004dy} B.~B.~Back {\it et al.}, Phys.\ Rev.\ C {\bf 70}, 021902(R) (2004). 
\bibitem{Back:2003xk} B.~B.~Back {\it et al.}, arXiv:nucl-ex/0301017.
\bibitem{Back:2003sr} B.~B.~Back {\it et al.}, Nucl.\ Instrum.\ Meth.\ A {\bf 499}, 603 (2003).
\bibitem{Gyulassy:1994ew} M.~Gyulassy and X.~N.~Wang, Comput.\ Phys.\ Commun.\  {\bf 83}, 307 (1994).
\bibitem{Back:2004je} B.~B.~Back {\it et al.}, Nucl. Phys. A. 757, 28 (2005).
\bibitem{ua5} G.~J.~Alner {\it et al.}, Z.\ Phys.\ C {\bf 33}, 1 (1986).
\bibitem{ua5physrep} H.~Heiselberg, Phys.\ Rept.\  {\bf 351}, 161 (2001).
\bibitem{Abe:1989td} F.~Abe {\it et al.}, Phys.\ Rev.\ D {\bf 41}, 2330 (1990).
\bibitem{Back:2004mr} B.~B.~Back {\it et al.}, nucl-ex/0409021, submitted to Phys. Rev. C (in press, 2005).
\bibitem{Back:2004zg} B.~B.~Back {\it et al.}, Phys.\ Rev.\ Lett.\  {\bf 94}, 122303 (2005).
\bibitem{Back:2004ra} B.~B.~Back {\it et al.}, Phys.\ Rev.\ Lett.\  {\bf 94}, 082304 (2005).
\bibitem{Busza:1991ah} W.~Busza, Nucl.\ Phys.\ A {\bf 544}, 49 (1992).
\bibitem{elias} J.~E.~Elias {\it et al.} Phys.\ Rev.\ Lett.\  {\bf 41}, 285 (1978).
\bibitem{wounded} A.~Bia\l as, B.~Bleszy\'{n}ski and W.~Czy\.{z}, Nucl.\ Phys.\ {\bf B111}, 461 (1976).
\end{references}
\end{document}